\documentclass[aps,prd,twocolumn,superscriptaddress,nofootinbib]{revtex4}
\usepackage{epsfig,epsf}
\usepackage{amsmath}
\usepackage{amsthm}
\usepackage{amsfonts}
\usepackage{amssymb}
\usepackage{dsfont}
\usepackage{multirow}
\usepackage{appendix}
\usepackage{slashed}
\usepackage[active]{srcltx}
\usepackage{psfrag}
\usepackage[xspace]{ellipsis}

\begin{document}

\title{On the mass and decay constant of  the P-wave ground and radially excited $h_{c}$ and $h_{b}$ axial-vector mesons}
\date{\today}
\author{K.~Azizi}
\affiliation{School of Physics, Institute for Research in Fundamental Sciences (IPM),\\
P.O.Box 19395-5531, Tehran, Iran} \affiliation{Department of
Physics, Do\v{g}u\c{s} University, Acibadem-Kadik\"{o}y, 34722
Istanbul, Turkey}
\author{J.Y.~S\"ung\"u}
\affiliation{Department of Physics, Kocaeli University, 41380
Izmit, Turkey}

\begin{abstract}

The mass and decay constant of the heavy quarkonia $h_{Q}(1P)$ and
$h_{Q}(2P)$ (Q = b, c) with quantum numbers $J^{PC} =1^{+-}$ are calculated in the framework of the two-point QCD sum rule method
by taking into account the vacuum condensates up to eight dimensions.
We compare our results for parameters of the $h_{b}(1P)$ and
$h_{c}(1P)$ quarkonia, and their first radially excited states
$h_{b}(2P)$ and $h_{c}(2P)$ with available experimental data as well as
predictions of other theoretical studies existing in the
literature. The results of present work may shed light on experimental searches for the $h_{c}(2P)$ state.
\end{abstract}

\maketitle

\section{Introduction}

It is well-known that the heavy quarkonia can take over the same inscription in probing  the QCD as
the hydrogen atom did in the atomic physics. The search for
heavy flavor production especially the quarkonia has  played a prominent role and
offered an insight into the dynamics of the strong interaction. A reliable description of heavy quarkonium states is of
great interest not only for understanding their internal organizations but for our knowledge of non-perturbative aspects of
QCD. Research on bottomonium and
charmonium states can  provide constraints on models of
quarkonium spectroscopy, as well
(for more information on the importance of quarkonia see for instance \cite{Cho:1995,Satz:2005,Rapp:2008,Voloshin:2007,Beringer:2012,Brambilla:2010,Soto:2017,YWang:2017,DeMori:2017,Ebert:2005jj}).

The  production mechanism of quarkonium states in the experiment
has been challenging for a long time and has presented important
disagreements with the theoretical predictions
\cite{Eichten:1979,Cacciari:1999,Fulsom:2014iha}. To overcome this
problem physicists  have come to the idea that there might be a
new set of particles. As a result of these efforts many XYZ states have been discovered \cite{Aubert:2005rm,Choi:2007wga,Mizuk:2008,Krokovny:2013mka,Vinodkumar:2012tzs,D0:X5568,X3872disc}. These states were then interpreted as new multiquark configurations not fitting to the standard quark model. 
These developments have expanded our knowledge on hadrons and triggered wide
discussions ended up in a new research area. Yet, unfortunately no
a complete  theoretical model has been established  that could
have a global description of what has been observed. New surveys
with more production and decay mechanisms and search for possible
partners having similar configurations may provide us with useful
knowledge on their internal structures, quark organizations  and
interactions \cite{Andronic:2015wma}. There have been  a lot of
experimental attempts on the  spectroscopy of the new states in
vacuum by different collaborations such as  CLEO, LHCb, Belle,
BESIII, etc.  The PANDA experiment at FAIR is planned to begin
taking data in 2025 aiming to explore the properties of
charmonium-like particles in details.

Despite the impressive developments in the experimental side, not all of theoretically predicted charmonium states have
been found, and some $2P$ resonances in the
charmonium sector, such as the $h_c(2P)$ state are not described
well \cite{Eichten:2004,Barnes:2007}. A clear identification
of the mass of $h_c(2P)$ meson would complete all charmonium
states expected by quark models to fall below the $D\overline{D}$
threshold (see also \cite{Rosner:2011sj}).

The P-wave state $h_{c}(1P)$ has been confirmed frequently
since its first detection in the $p\bar{p}$ collisions by the
R704 collaboration \cite{R704exp}. The FNAL E760 experiment looked
for the $h_{c}$ in the reaction $p\bar{p}\rightarrow
h_c\rightarrow \pi^0 J/\psi\rightarrow e^+ e^-$ and announced a
statistically significant enhancement with
$M_{h_c}=3526.2\pm0.2\pm0.2\mathrm{MeV}$ and
$\Gamma_{h_c}\leq1.1\mathrm{MeV}$ \cite{E760exp}. The best
clue for the $h_{c}(1P)$ came out from the CLEO collaboration
when looking for the isospin violating transition $e^+e^- \rightarrow
\Upsilon(2S)\rightarrow h_c \pi^0$ \cite{CleoRosner}. The
mass of  $h_c(2^1P_1)$ was estimated around ($3934-3956) \mathrm{MeV}$
via HISH Model \cite{Son}. In 2016, the BESIII
collaboration reported a very good $h_{c}$ signal in the reaction
channels $h_{c}\rightarrow \gamma \eta'$ and $h_{c}\rightarrow
\gamma \eta$  with a statistical significance of
$8.4\sigma$ and $4.0\sigma$, respectively
\cite{BES3Ablikimradiative}. Very recently  two resonances
were observed in the $e^+e^- \rightarrow \pi^+\pi^- h_c$ process 
located in the mass region between $4.2\mathrm{GeV/c^2}$ and
$4.4\mathrm{GeV/c^2}$, where the vector charmonium hybrid states
have been predicted by various QCD calculations
\cite{Ablikimhibrit2017}.

As for the spin-singlet P-wave bottomonium state $h_{b}(1P)$, the
BaBar collaboration first reported the evidence in the sequential
process $\Upsilon(3S)\rightarrow \pi^{0}h_{b}(1P)
\rightarrow\pi^{0}\gamma\eta_{b}(1S)$ in 2011
\cite{Lees:2011Babar}. The resonance $h_{b}(1P)$ was investigated
and detected later again by Belle in the transition
$\Upsilon(5S)\rightarrow h_{b}(1P)\pi^{+}\pi^{-}$
\cite{Bondar:Belle2012,Adachi:Belle}.

The P-wave singlet charmonium and bottonium systems  can give precious clues on spin-spin (or
hyperfine) interactions between quarks at which the complications due to
the relativistic effects are less important compared to the light quark
systems. The way of analyzing the spin-spin interactions between quarks
is to introduce a central potential term. In a QCD-based potential
model, the potential term is commonly described as a Coulomb term
stemming from one-gluon exchange plus a confining term. Regarding
this potential, the hyperfine splitting between the spin-singlet
and spin-triplet P-wave states  is  very small \cite{Buchmuller:2012}.

For instance, the $Y(4260)$ resonance has been considered in
literature as a hybrid with a color octet $c\bar{c}$ pair bound
with a gluon. However, a new decay is observed,
$Y(4260)\rightarrow h_{c}\pi^{+} \pi^{-}$
\cite{Ablikimhibrit2017}, which would imply a spin flip of the
heavy quark system. If another measurement will confirm this
production mechanism, the hybrid interpretation of the $Y(4260)$ state
would be potently disregarded. A similar case is found in the
decay channel $\Upsilon(10860)\rightarrow
\Upsilon(1S,2S,3S)\pi^{+} \pi^{-}$ and $\Upsilon(10860)\rightarrow
h_{b}(1P,2P)\pi^{+} \pi^{-}$. Meanwhile, discoveries of the two
charged bottomonia states $Z^{+}_b(10610)$ and $Z^{+}_b(10650)$
\cite{Adachi:Belle}, found in the decays $\Upsilon(10860)/Y_b
\rightarrow Z^{+}_b(10610) \pi^{-}, Z^{+}_b(10650) \pi^{-}$,
yielding  $\Upsilon(1S)\pi^+ \pi^-,
\Upsilon(2S)\rightarrow \pi^+ \pi^-, \Upsilon(3S)\pi^+\pi^-,
h_b(1P)\pi^+ \pi^-$ and $h_b(2P)\pi^+ \pi^-$ final states, indicates a
tetraquark interpretation. In the tetraquark picture, the two
$Z_b$ states have both (spin-0) and (spin-1) components in Fock
space. Finally decays of some exotica into $h_b$ and $h_c$ mesons
is very crucial for determining their inner structures and
dynamics \cite{AhmetAli2017}.

The  mesons $h_{c}(1P)$ and $h_{b}(1P)$ have been previously
investigated using different models such as Extended Potential
Model \cite{Akbar:2015EPM}, Nonrelativistic Quark Models
\cite{Deng:2016QM}, Friedrichs-model-like Scheme
\cite{Zhou:2017FMLS}, Relativistic Quark Model \cite{Bhat:2017RQM,
Souza:2017RQM}, Quark Model \cite{Godfrey:1985dt}, Screened
potential \cite{Li:2009zu},  Holography Inspired Stringy
Hadron(HISH) \cite{Son} and Lattice QCD \cite{Becirevic:2013bsa}.

In the present work, we study the ground-state heavy quarkonia
$h_{b}(1P)$ and $h_{c}(1P)$, and their first radially excited
states $h_{b}(2P)$, and $h_{c}(2P)$ via the QCD sum rule method
\cite{Shifman1,Shifman2}, and make predictions for their masses
and decay constants. It is known that QCD sum rule is one of the
powerful and effective non-perturbative methods that provide
valuable information in the search for the excited quarkonium
states and exotica. We compare also our results with
 available experimental data and relevant theoretical predictions
presented in the literature.

The paper is arranged in the following way. In section II we
briefly review the theoretical background for QCD sum rules and present
 details of the mass and decay constant calculations for 
hidden-charm and bottom states with $J^{PC}=1^{+-}$.
In section III we discuss the results and compare our conclusions
 with ones obtained in the context of
other models. Last section is devoted to the summary and outlook.

\section{Determination of Mass and Decay Constant of $h_{c}$ and $h_{b}$ States}

The aim of QCD sum rule method is to extract the hadronic observables
(e.g. masses, coupling constants etc.) from microscopic QCD
degrees of freedom  such as vacuum quark-gluon condensates and  quark masses. This
approach relates the micro world
of QCD at high energies to the hadronic sector at low energies.
The correlation function of two currents is introduced and treated
by the help of the operator product expansion (OPE), where the short
and long-distance effects are separated. The
former is calculated with QCD perturbation theory, while the
latter is parameterized in terms of the quark-gluon vacuum condensates.

The strategy behind this technique is to interpret an appropriate correlation
function in two different ways. On  one hand it is identified with a hadronic propagator, called Physical
 (or Phenomenological) side. On the other hand, as previously mentioned,  the correlation function is calculated in terms of the quarks and gluons and their interactions with the QCD vacuum, called QCD (or
Theoretical) side. Then, the result of the QCD calculations is
matched to a sum over the hadronic states via a dispersion
relation. The sum rules obtained  allow one to calculate different
observable characteristics of the hadronic ground and excited
states. The interpolating currents representing the hadronic
states in this approach couple not only to their ground states,
but also to their  excited states with the same quark contents and
quantum numbers. Therefore, via this method, in addition to the
parameters such as the mass and decay constant of the ground state
$h_{Q}(1P)$, spectroscopic parameters of its first radial
excitation, i.e. $h_{Q}(2P)$ can be calculated, as well.

In this section, we present the details of the calculation of the
masses and decay constants of the $h_{b}(1P),h_{b}(2P),h_{c}(1P)$
and $h_{c}(2P)$ mesons with $J^{PC}=1^{+-}$.    The starting point is
to deal with the following two-point correlation function
\cite{Shifman1,Shifman2}:
\begin{equation}\label{eq:CorrFunc}
\Pi_{\mu \nu }(p)=i\int d^{4}x~e^{ip\cdot x}\langle
0|\mathcal{T}\{J_{\mu}^{h_Q}(x)J_{\nu}^{\dagger h_Q}(0)\}|0\rangle,
\end{equation}
where $J_{\mu}(x)$ is the interpolating current with the quantum
numbers $J^{PC}=1^{+-}$\cite{Reinders}:
\begin{equation}\label{eq:Current}
J_{\mu}^{h_{Q}}(x)=\bar{Q}_{i}(x)\vec{\partial}_{\mu}(x)\gamma_{5}{Q}_{i}(x),
\end{equation}
where $Q=b$ or  $c$, and $i$ is the color index.

Firstly, we determine the sum rules for the masses $m_{h_Q}$ and
decay constants $f_{h_Q}$ of the ground states. The decay constant of a state represents the relation of the hadronic state with the vacuum through its interpolating current. This is the main input in analyses of the possible strong, electromagnetic and weak decays of hadrons in order to estimate their total width.  To calculate the parameters of the ground states, we
employ the ``ground state + continuum'' approximation. Later the
``ground state +first excited state + continuum'' approximation is
used to derive sum rules for the $h_{b}(2P)$ and $h_{c}(2P)$
mesons. So the masses and decay constants of $h_{c}(2P)$ and
$h_{b}(2P)$ can be extracted from these expressions. Obtained
numerical values for the parameters of the ground states are used as inputs in the sum rules for the excited $h_{b}(2P)$ and
$h_{c}(2P)$ mesons.

To obtain the physical side, a complete set of intermediate
hadronic states with the same quantum numbers as the current
operator $J_{\mu}(x)$ can be inserted into the correlation
function. Then isolating the terms that we are interested in from other
quarkonium states and carrying out the integration over $x$, we
obtain the following expression:
\begin{eqnarray}
&&\Pi _{\mu \nu }^{\mathrm{Phys.}}(p)=\frac{\langle 0|J_{\mu
}^{h_{Q}(1P)}|h_{Q}(1P)\rangle \langle h_{Q}(1P)|J_{\nu }^{\dagger h_{Q}(1P)}|0\rangle}{%
m_{h_{Q}(1P)}^{2}-p^{2}}  \notag \\
&&+ \frac{\langle 0|J_{\mu }^{h_{Q}(2P)}|h_{Q}(2P)\rangle \langle
h_{Q}(2P)|J_{\nu }^{h_{Q}(2P)\dagger}|0\rangle
}{m_{h_{Q}(2P)}^{2}-p^{2}}+\ldots,
\label{eq:Phys1}
\end{eqnarray}
where $m_{h_{Q}(1P)}$ and $m_{h_{Q}(2P)}$ are the masses of
$h_{Q}(1P)$ and $h_{Q}(2P)$ states, respectively. The  ellipsis  in
Eq.\ (\ref{eq:Phys1})  represent contributions coming from higher
resonances and continuum states.

To complete the calculations of the phenomenological side of the
sum rules we introduce the matrix elements through masses and
decay constants of $h_{Q}(1P)$ and $h_{Q}(2P)$ mesons as
\begin{eqnarray}
&&\langle0|J_{\mu}^{h_{Q}(1P)}|h_{Q}(1P)\rangle=f_{h_{Q}(1P)}m^2_{h_{Q}(1P)}\varepsilon_{\mu}, \nonumber\\
&&\langle0|J_{\mu}^{h_{Q}(2P)}|h_{Q}(2P)\rangle=f_{h_{Q}(2P)}m^2_{h_{Q}(2P)}\tilde{\varepsilon}_{\mu}
\label{eq:CurDef.}
\end{eqnarray}
where $\varepsilon _{\mu }$ and $\tilde{\varepsilon}_{\mu }$ are
the polarization vectors of the $h_{Q}(1P)$ and $h_{Q}(2P)$
states, respectively. So the function $\Pi _{\mu \nu
}^{\mathrm{Phys.}}(p)$ can be written as
\begin{eqnarray}
&&\Pi _{\mu \nu
}^{\mathrm{Phys.}}(p)=\frac{m_{h_{Q}(1P)}^{4}f_{h_{Q}(1P)}^{2}}{
m_{h_{Q}(1P)}^{2}-p^{2}}\Bigg( -g_{\mu \nu }+\frac{p_{\mu }p_{\nu
}}{
m_{h_{Q}(1P)}^{2}}\Bigg)  \notag \\
&&+\frac{m_{h_{Q}(2P)}^{4}f_{h_{Q}(2P)}^{2}}{m_{h_{Q}(2P)}^{2}-p^{2}}\Bigg(-g_{\mu \nu }+\frac{%
p_{\mu }p_{\nu }}{m_{h_{Q}(2P)}^{2}}\Bigg) +\ldots.
\label{eq:Phys2}
\end{eqnarray}
Then Borel transformation applied to Eq.\ (\ref{eq:Phys2}) yields
\begin{eqnarray}
\mathcal{B}(p^2)~\Pi_{\mu\nu}^{\mathrm{Phys.}}(p)&=&m_{h_{Q}(1P)}^{4}f_{h_{Q}(1P)}^{2}e^{-m_{h_{Q}(1P)}^{2}/{M}^{2}}\notag \\
&\times&\left(-g_{\mu\nu}+\frac{p_{\mu}p_{\nu}}{m_{h_{Q}(1P)}^{2}}\right)  \notag \\
&+&m_{h_{Q}(2P)}^{4}f_{h_{Q}(2P)}^{2}e^{-m_{h_{Q}(2P)}^{2}/{M}^{2}}\notag \\
&\times&\left(-g_{\mu\nu}+\frac{p_{\mu}p_{\nu}}{m_{h_{Q}(2P)}^{2}}\right)+\ldots,
\label{eq:Phys3}
\end{eqnarray}
where $M^{2}$ is the Borel parameter that should be fixed.

The correlation function on QCD side,  $\Pi_{\mu\nu
}^{\mathrm{QCD}}(p)$, can be written down by  contracting the
heavy quark fields. After simple manipulations and putting $y=0$,
it reads
\begin{eqnarray} \label{eq:CorrFunc2}
\Pi _{\mu \nu }^{\mathrm{QCD}}(p)&=&i\int d^{4}x~e^{ip.x} \notag \\
&\times&\mathrm{Tr}~[\vec{\partial}_{\mu}(x)
\vec{\partial}_{\nu}(y){S}_{Q}^{ji}(y-x)\gamma_5
S_{Q}^{ij}(x-y)\gamma _{5}]\vert_{y=0},\nonumber\\
\end{eqnarray}
 where the $S_Q^{ij}$ is the
heavy quark propagator explicit form of which is presented below
\cite{Reinders};
\begin{eqnarray}
&&S_{Q}^{ij}(x)=i\int \frac{d^{4}k}{(2\pi )^{4}}e^{-ikx}\Bigg
\{\frac{\delta
_{ij}\left( {\slashed k}+m_{Q}\right) }{k^{2}-m_{Q}^{2}}  \notag \\
&&-\frac{g_{s}G_{ij}^{\alpha \beta }}{4}\frac{\sigma _{\alpha \beta }\left( {%
\slashed k}+m_{Q}\right) +\left( {\slashed k}+m_{Q}\right) \sigma
_{\alpha
\beta }}{(k^{2}-m_{Q}^{2})^{2}}  \notag \\
&&+\frac{g_{s}^{2}G^{2}}{12}~\delta _{ij}~m_{Q}\frac{k^{2}+m_{Q}{\slashed k}}{%
(k^{2}-m_{Q}^{2})^{4}}+\frac{g_{s}^{3}G^{3}}{48}~\delta _{ij}\frac{\left( {%
\slashed k}+m_{Q}\right) }{(k^{2}-m_{Q}^{2})^{6}}  \notag \\
&&\times \left[ {\slashed k}\left( k^{2}-3m_{Q}^{2}\right)
+2m_{Q}\left( 2k^{2}-m_{Q}^{2}\right) \right] \left( {\slashed
k}+m_{Q}\right)+...\Bigg \},  \notag \\
\label{eq:Qprop}
\end{eqnarray}
In Eq.\ (\ref{eq:Qprop}) we used the following notations
\begin{eqnarray}
&&G_{ij}^{\alpha \beta }=G_{A}^{\alpha \beta
}t_{ij}^{A},\,\,~~G^{2}=G_{\alpha \beta }^{A}G_{\alpha \beta
}^{A},  \notag\\
&&G^{3}=\,\,f^{ABC}G_{\mu \nu }^{A}G_{\nu \delta }^{B}G_{\delta
\mu }^{C}, \label{eq:A.3}
\end{eqnarray}
with $i,\,j=1,2,3$ being the color indices  and $A,B,C=1,\,2\,\ldots 8$. In Eq.\ (\ref{eq:A.3}), $t^{A}=\lambda ^{A}/2$,
$\lambda ^{A}$ are the Gell-Mann matrices and the gluon field
strength tensor $G_{\alpha \beta }^{A}\equiv G_{\alpha \beta
}^{A}(0)$ is fixed at $x=0$.

The function $\Pi _{\mu \nu }^{\mathrm{QCD}}(p)$ has two different
structures, and can be expressed as a sum of two components as
follows
\begin{equation}
\Pi _{\mu \nu }^{\mathrm{QCD}}(p)=\Pi ^{\mathrm{QCD}}(p^{2})(-g_{\mu \nu })+%
\widetilde{\Pi }^{\mathrm{QCD}}(p^{2})p_{\mu }p_{\nu }.
\end{equation}
The QCD sum rules for the physical quantities of $h_{Q}(2P)$ can
be extracted after equating the coefficient of the same structures in both $\Pi _{\mu \nu }^{\mathrm{Phys}%
}(p)$ and $\Pi _{\mu \nu }^{\mathrm{QCD}}(p)$. To continue our
evaluations, we select the structure $p_{\mu} p_{\nu}$. The
invariant function $\widetilde{\Pi } ^{\mathrm{QCD}}(p^{2})$  corresponding to
this structure can be represented as the dispersion integral
\begin{equation}
\widetilde{\Pi }^{\mathrm{QCD}}(p^{2})=\int_{4m_{Q}^{2}}^{\infty}ds~\frac{\rho
^{\mathrm{QCD}}(s)}{s-p^{2}}+\mathrm{subtracted~terms},
\end{equation}
where $\rho^{\mathrm{QCD}}(s)$ is the corresponding  two-point spectral density.
It consists of two parts and can be expressed as
\begin{equation}
\rho^{\mathrm{QCD}}(s)=\rho^\mathrm{Pert.}(s)+\rho^\mathrm{Nonpert.}(s).
\label{eq:rho}
\end{equation}
Here $\rho^{\mathrm{Pert.}}(s)$ is the  perturbative part of the
spectral density that is given by the formula
\begin{equation}
\rho^{\mathrm{Pert.}}(s)=\frac{1}{8\pi^2}(-12m_Q^2+3s).
\label{eq:rhopert}
\end{equation}
The non-perturbative part of
 the spectral density is presented in the Appendix. After
applying the Borel transformation  on the variable $p^{2}$ to both
the physical and QCD sides of the sum rules and subtraction of the
contributions of the higher states and continuum  using
 the quark-hadron duality assumption, we find the required sum rules. The
sum rules for the mass and decay constant of the excited
$h_{Q}(2P)$ states in terms  of the parameters of the $h_{Q}(1P)$ mesons, are found as:

\begin{widetext}

\begin{equation}
m^{2}_{h_{Q}(2P)}=\frac{\int_{4m_{Q}^{2}}^{s_{0}^{\star}}ds~\rho^{\mathrm{QCD}
}(s)~s~e^{-s/M^{2}}-f_{h_{Q}(1P)}^{2}m_{h_{Q}(1P)}^{4}e^{-m_{h_{Q}(1P)}^{2}/M^{2}}}{\int_{4m_{h_{Q}}^{2}}^{s_{0}^{\star
}}ds~\rho^{\mathrm{QCD}}(s)~e^{-s/M^{2}}-f_{h_{Q}(1P)}^{2}m_{h_{Q}(1P)}^{2}e^{-m_{h_{Q}(1P)}^{2}/M^{2}}},
\label{eq:MassSR}
\end{equation}
and
\begin{eqnarray}
f_{h_{Q}(2P)}^{2}=\frac{1}{m_{h_{Q}(2P)}^{2}}\Bigg[
\int_{4m_{Q}^{2}}^{s_{0}^{\star}}ds~\rho
^{\mathrm{QCD}}(s)~e^{\Big(m_{h_{Q}(2
P)}^{2}-s\Big)/M^{2}}
-f_{h_{Q}(1P)}^{2}m_{h_{Q}(1P)}^{2}e^{\Big(m_{h_{Q}(2P)}^{2}-m_{h_{Q}(1P)}^{2}\Big)/M^{2}}\Bigg].
\label{eq:DecConSR}
\end{eqnarray}

\end{widetext}

Here $s_{0}^{\star }$ is the continuum threshold parameter
separating the contribution of the ``$h_{Q}(1P)+h_{Q}(2P)$'' states
from the contribution due to ``higher resonances and continuum''.
As we previously mentioned the mass and decay constant of
$h_{Q}(1P)$ enter into Eqs.\ (\ref{eq:MassSR}) and
(\ref{eq:DecConSR}) as the inputs. The mass of the
$h_{Q}(1P)$ state can be extracted from the sum rule
\begin{equation}
m_{h_{Q}(1P)}^{2}=\frac{\int_{4m_{Q}^{2}}^{s_{0}}ds~\rho ^{\mathrm{QCD}%
}(s)~s~e^{-s/M^{2}}}{\int_{4m_{Q}^{2}}^{s_{0}}ds~\rho^{%
\mathrm{QCD}}(s)~e^{-s/M^{2}}}, \label{eq:MassSR1P}
\end{equation}
whereas to obtain the numerical value of the decay constant
$f_{h_{Q}(1P)}$ we use the following expression
\begin{equation}
f_{h_{Q}(1P)}^{2}=\frac{1}{m_{h_{Q}(1P)}^{2}}\int_{4m_{Q}^{2}}^{s_{0}}ds~\rho
^{\mathrm{QCD}}(s)e^{\Big(m_{h_{Q}(1P)}^{2}-s\Big)/M^{2}}.
\label{eq:DecConSR1P}
\end{equation}
In Eqs.\ (\ref{eq:MassSR1P}) and  (\ref{eq:DecConSR1P}) $s_0$ is
the continuum threshold, which separates the contribution of the
ground state $h_{Q}(1P)$  from those of the higher resonances and continuum. The sum rules for the ground and first radially excited
 states  contain the same spectral density $\rho
^{\mathrm{QCD}}(s)$, but the continuum threshold has to obey $s_0
< s_0^{\star}$.

\section{Numerical Analysis}

The sum rules obtained in this study allow  us to calculate
characteristics of the  ground-state mesons and their first radial
excitations. The obtained sum rules depend on the
Borel mass parameter $M^2$ and continuum threshold $s_0$.
Nevertheless, the dependence of physical quantities extracted from the sum
rules on these auxiliary parameters should remain inside the standard  limits, allowed by the method used, i. e.  the uncertainties should not exceed the $ 30\% $ of the total values. These limits are  determined by the systematic errors of the method coming from the quark-hadron duality assumption and those belong to the variations of the auxiliary parameters as well as other inputs.
The sum rules found  include the vacuum
expectations of the different gluon operators as well as the heavy quark masses as input parameters, numerical values of which are  presented in Table I.

\begin{table}[tbp]
\begin{tabular}{|c|c|}
\hline\hline Parameters & Values  \\
\hline\hline
$m_{c}$ & $(1.67\pm0.07)~\mathrm{GeV}$ \\
$m_{b}$ & $(4.78\pm0.06)~\mathrm{GeV}$ \\
$\langle\frac{\alpha_sG^2}{\pi}\rangle $ & $(0.012\pm0.004)~\mathrm{GeV}%
^4 $\\
$\langle g_{s}^3G^3\rangle$ & $(0.57\pm0.29)~\mathrm{GeV}^6 $\\
\hline\hline
\end{tabular}
\caption{Input parameters (see, Refs.\
\cite{Patrignani:2016xqp,Shifman1,Shifman2,Ioffe:2005ym,Belyaev:1982cd}).}
\label{tab:Param}
\end{table}
\begin{table}[tbp]
\begin{tabular}{|c|c|c|}
\hline\hline Resonance & $h_{c}$ & $h_{b}$\\\hline\hline
$M^2~(\mathrm{GeV}^2$) & $3-6$ & $10-16$\\\hline
$s_0 ~(\mathrm{GeV}^2$) &$13-15$ & $100-104$\\
$s_0^{\star}~(\mathrm{GeV}^2$) &$16-18$ & $107-111$\\

\hline\hline
\end{tabular}%
\caption{Values of the Borel parameter and continuum thresholds
used in this study to evaluate parameters of the  $h_{c}$ and
$h_{b}$ mesons.} \label{tab:Msqs0values}
\end{table}

The numerical analyses performed  allow us to fix the working intervals  of the parameters
$M^2$ and $s_0$, where the standard conditions (pole dominance and OPE convergence) are
satisfied. The upper bound on  $M^{2}$ is found requiring that  the contributions of the resonances under consideration exceed the contributions of the higher states and continuum. Its lower bound is found demanding the convergence of the OPE and exceeding the perturbative part over the total non-perturbative contributions. The parameters $s_0$ and $s_{0}^{*}$,  are determined from
the conditions that guarantee the sum rules to have the best
stability in the allowed $M^{2}$ regions. This is possible by achieving the maximum pole contributions and the best convergence of the OPE.
  The obtained working
region for the Borel parameter and continuum thresholds are
presented in Table \ref{tab:Msqs0values}.

In figs. (\ref{fig:Masshc1P2P})- (\ref{fig:DecConhb2P}), we show the dependence of
$m_{h_{Q}(1P)}$, $f_{h_{Q}(1P)}$, $m_{h_{Q}(2P)}$ and
$f_{h_{Q}(2P)}$ on $M^2$ at fixed $s_0$, and as functions of $s_0$
for chosen values of $M^2$. The masses of the $h_{Q}$
mesons are rather stable with respect to the  variations  of the auxiliary parameters $M^2$ and
$s_{0}^{(*)}$, compared to the decay constants $f_{h_Q}(1P,2P)$ which are relatively sensitive  to the changes of the
auxiliary parameters. The logic behind this is rather simple:
 the sum rules for the masses of the states under consideration are
obtained as ratios of two integrals over the spectral densities $\rho(s)$ and $
s\rho(s)$, which considerably decrease the  effects due to the changes of the $M^2$ and $
s_0$. On the contrary, the decay constants depend on the aforesaid
integrals themselves, and therefore,  undergone relatively sizable
variations. Nevertheless, theoretical errors for $f_{h_{Q}(1P)}$ and $f_{h_{Q}(2P)}$ arising from uncertainties of $
M^2$ and $s_0$, and other input parameters stay within the allowed
intervals for the theoretical uncertainties ingrained in sum rule calculations
which may reach to, as we previously mentioned,  $30\%$ of the total values.

The numerical values extracted from the sum rules for the physical quantities are collected in Table
\ref{tab:MassResults}, where we write down the
masses of the mesons $h_{Q}(2P)$ and $
h_{Q}(1P)$. The presented errors belong to the variations of the results with respect to the variations of the auxiliary parameters in their working region as well as those uncertainties coming from the other input parameters.   We  compare our  predictions with the existing experimental data as well as other theoretical results. It is seen, that $m_{h_{c}(1P)}$, $m_{h_{b}(1P)}$ and $m_{h_{b}(2P)}$
are  in agreement with the existing experimental data within the
errors. The results of all theoretical studies on the masses of the states under consideration are roughly in agreements  with each other within the uncertainties.

Our results for the decay constants of corresponding mesons compared to other theoretical predictions are
presented in Table  \ref{tab:DecConResults}. We see overall considerable differences among different theoretical predictions on the decay constants of the states under consideration. Some predictions are consistent with each other within the errors. Even if we consider the errors of some theoretical predictions on the decay constants,  however, the results differ up to a factor of two. The decay constants are the main inputs in the calculations of the total widths of the states under considerations via their possible strong, electromagnetic and weak decays. Further experimental data on the width and mass of the considered states are needed in order to determine which of the  presented nonperturbative methods does work, well. 
\begin{widetext}

\begin{table}[t]
\begin{tabular}{|c|c|c|c|c|}
\hline\hline Parameter $[\mathrm{MeV}]$& $h_c(1P)$ & $h_c(2P)$ & $h_b(1P)$ & $h_b(2P)$                                   \\
\hline\hline
$M_\mathrm{Exp.}$      &~~ $3525.38 \pm 0.11$ \cite{Patrignani:2016xqp}~~& - &~~ $9899.3 \pm  0.8$ \cite{Patrignani:2016xqp}~~&~~ $10259.8 \pm 1.2$ \cite{Patrignani:2016xqp}~~\\
\hline
$M_\mathrm{Our~Work}$  & $3581^{+61}_{-60}$ & $3897^{+68}_{-69}$ & $9854^{+61}_{-57} $ & $10267^{+61}_{-68}$   \\
\hline
$M_\mathrm{Other~T.W.}$& 3516 \cite{Barnes:2005}    & 3934 \cite{Barnes:2005}   &9885 \cite{Ferretti:2013vua}& 10247 \cite{Ferretti:2013vua} \\
                       & 3517 \cite{Barnes:2005}    & 3956 \cite{Barnes:2005}   &9882 \cite{Godfrey:2015dia} & 10250 \cite{Godfrey:2015dia} \\
                       & 3525 \cite{Ebert:2011jc}   & 3926 \cite{Ebert:2011jc}  &9900 \cite{Ebert:2011jc}    & 10260 \cite{Ebert:2011jc}  \\
                       & 3522 \cite{Godfrey:2015dia}& 3955 \cite{Souza:2017RQM} &9879 \cite{Segovia:2016xqb} & 10240 \cite{Segovia:2016xqb} \\
                       &   &   &9915.5 \cite{Liu:2011yp}    & 10259.1 \cite{Liu:2011yp}    \\
                       & 3519 \cite{Li:2009zu}      & 3908 \cite{Li:2009zu}     &9884.4 \cite{Lu:2016mbb}    & 10262.7 \cite{Lu:2016mbb}   \\
                       &                            & 3902 \cite{Zhou:2017dwj} &9898.95 \cite{Bhat:2017RQM} &  \\
                       & $3474 \pm 20$ \cite{Chen:2000ej} & $3886 \pm 92$ \cite{Chen:2000ej} &$9940 \pm 37$ \cite{Bashiry:2011} &$10269.15$ \cite{Bhat:2017RQM}  \\
                       & $3474 \pm 40$ \cite{Okamoto:2001jb}& $4053 \pm 95$ \cite{Okamoto:2001jb}&$9886^{+81}_{-78}$ \cite{AgaevZb}& $10331^{+108}_{-117}$ \cite{AgaevZb} \\
\hline\hline
\end{tabular}
\caption{The numerical values of the $h_{Q}(1P)$ and $h_{Q}(2P)$
mesons' masses.} \label{tab:MassResults}
\end{table}

\begin{table}[t]
\begin{tabular}{|c|c|c|c|c|}\hline\hline
Parameter & $h_c(1P)$ & $h_c(2P)$ & $h_b(1P)$ & $h_b(2P)$ \\
\hline\hline
$f_\mathrm{Our~Work}[\mathrm{MeV}]$   & $176^{+35}_{-35}$ & $244^{+37}_{-38}$ &$293^{+42}_{-41}$  &$318^{+52}_{-55}$ \\
\hline
$f_\mathrm{Other~T.W.}[\mathrm{MeV}]$ & 206  \cite{GLWang:2007} & 207 \cite{GLWang:2007} &  129 \cite{GLWang:2007} & 131 \cite{GLWang:2007} \\
                                      & 335 \cite{Novikov:1977dq} &        & $325^{+61}_{-57}$ \cite{AgaevZb} & $286^{+58}_{-53}$ \cite{AgaevZb} \\
                                      & $340^{+119}_{-101}$ \cite{hcDecConQM} &     & $94 \pm 10$ \cite{Bashiry:2011} &       \\
                                      & $490 \pm 2 \pm 40\pm 45$ \cite{Wang:2012gj}&  & $549 \pm 2\pm 50\pm 45$ \cite{Wang:2012gj}&  \\
                                      & $490 \pm 8\pm 40\pm 40\pm 44$ \cite{Wang:2012gj}&  & $552 \pm 3\pm 47\pm 46$ \cite{Wang:2012gj}&  \\

\hline\hline
\end{tabular}
\caption{The  decay constants of the $h_{Q}(1P)$ and $h_{Q}(2P)$
mesons.} \label{tab:DecConResults}
\end{table}

\begin{figure}[h!]
\begin{center}
\includegraphics[totalheight=6cm,width=8cm]{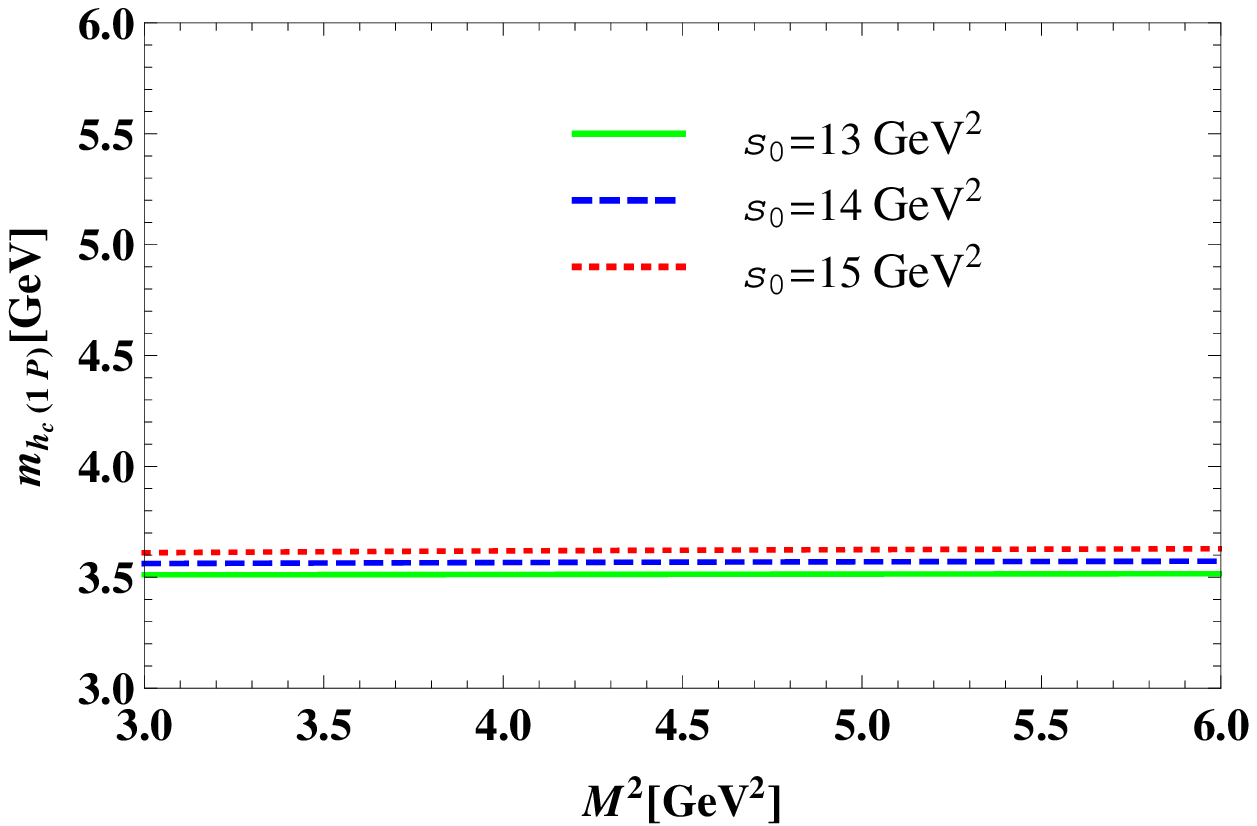}\,\,\,\,\,\,\,\,\,\,\,\,\,\,\,\,%
\includegraphics[totalheight=6cm,width=8cm]{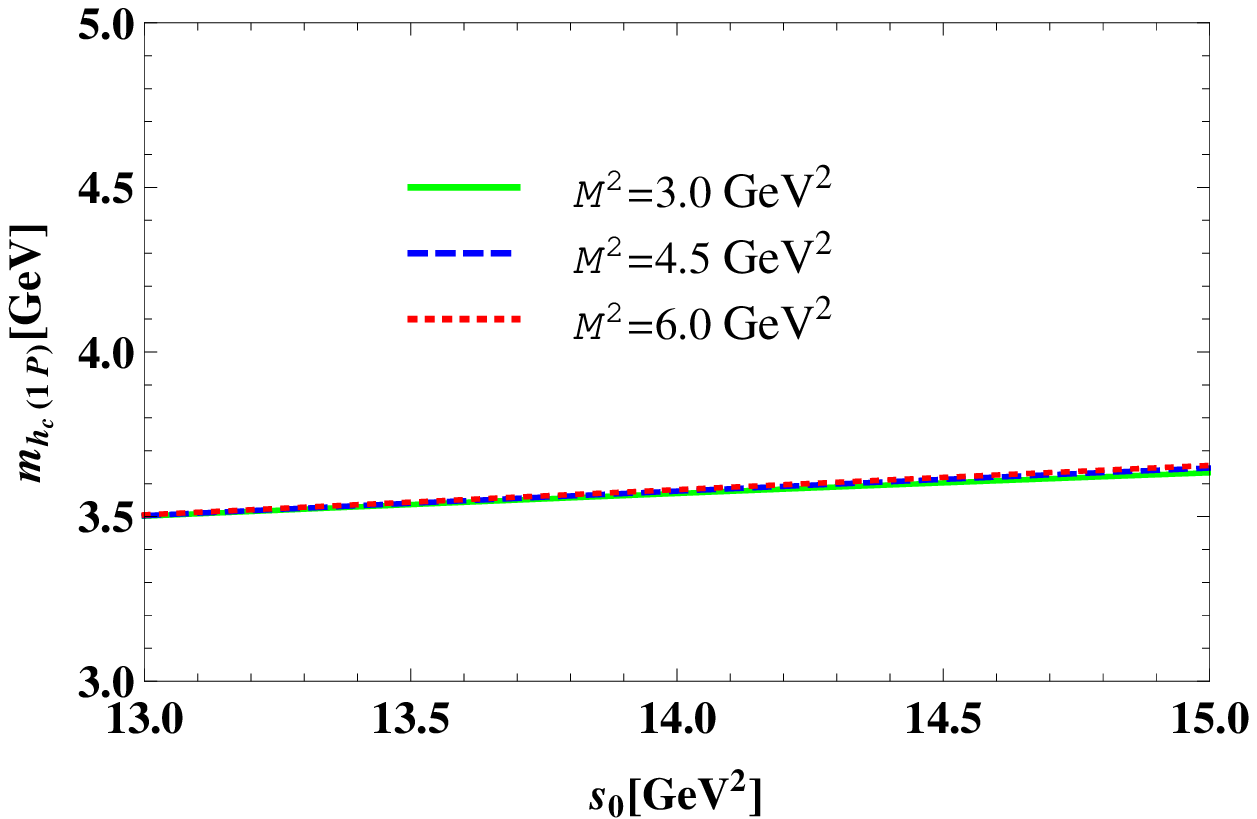}
\includegraphics[totalheight=6cm,width=8cm]{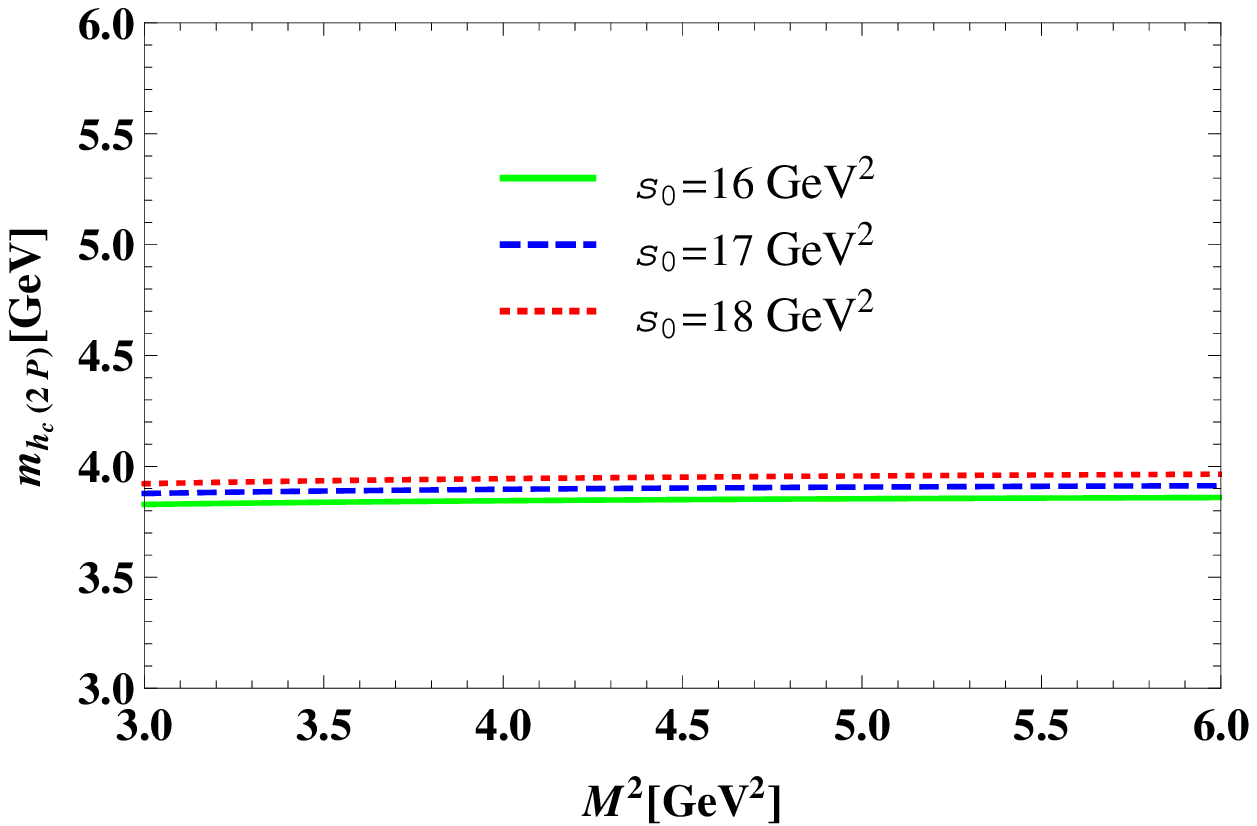}\,\,\,\,\,\,\,\,\,\,\,\,\,\,\,\,%
\includegraphics[totalheight=6cm,width=8cm]{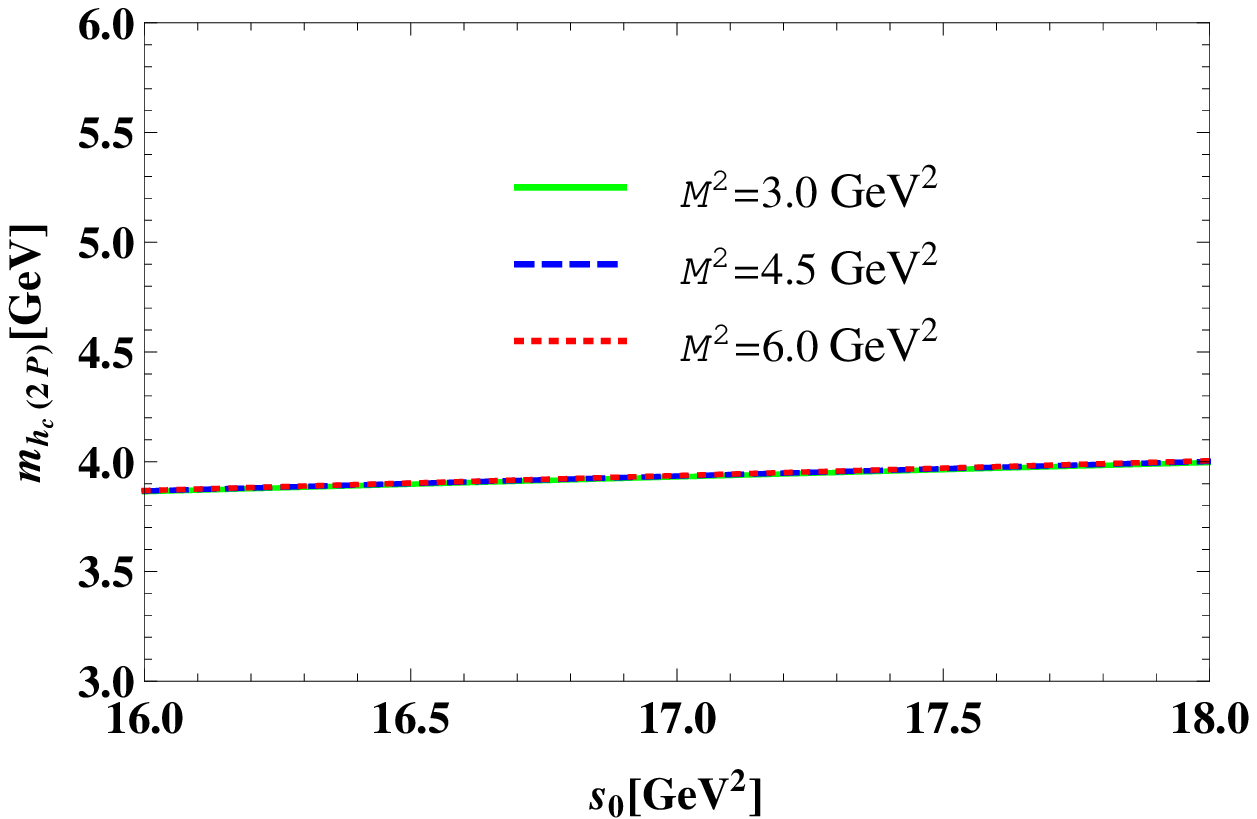}
\end{center}
\caption{ The masses of the mesons $h_c(1P,2P)$ in terms  of the
Borel parameter $M^2$ at fixed  values of $s_0$ (left panel), and in terms of the continuum threshold $s_0$ at fixed values of $M^2$ (right
panel).} \label{fig:Masshc1P2P}
\end{figure}

\begin{figure}[h!]
\begin{center}
\includegraphics[totalheight=6cm,width=8cm]{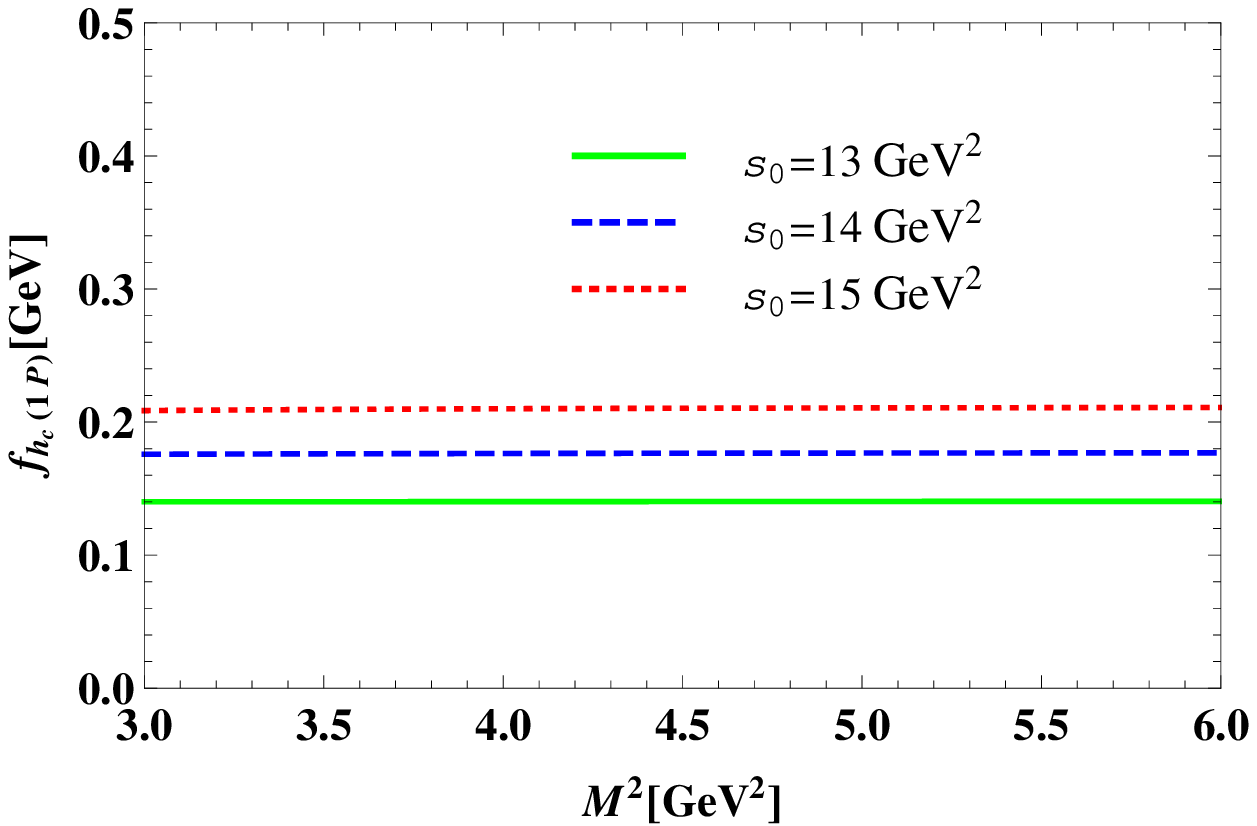}\,\,\,\,\,\,\,\,\,\,\,\,\,\,\,\,%
\includegraphics[totalheight=6cm,width=8cm]{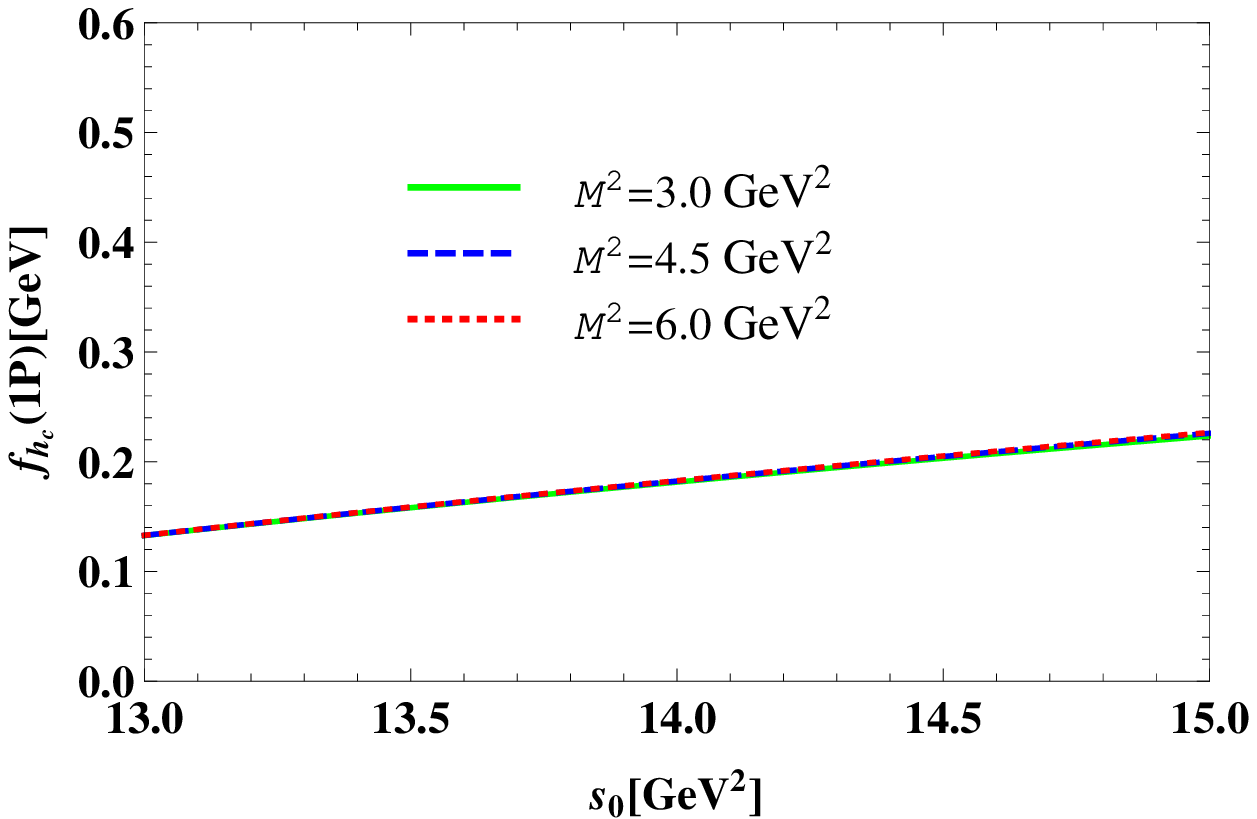}
\includegraphics[totalheight=6cm,width=8cm]{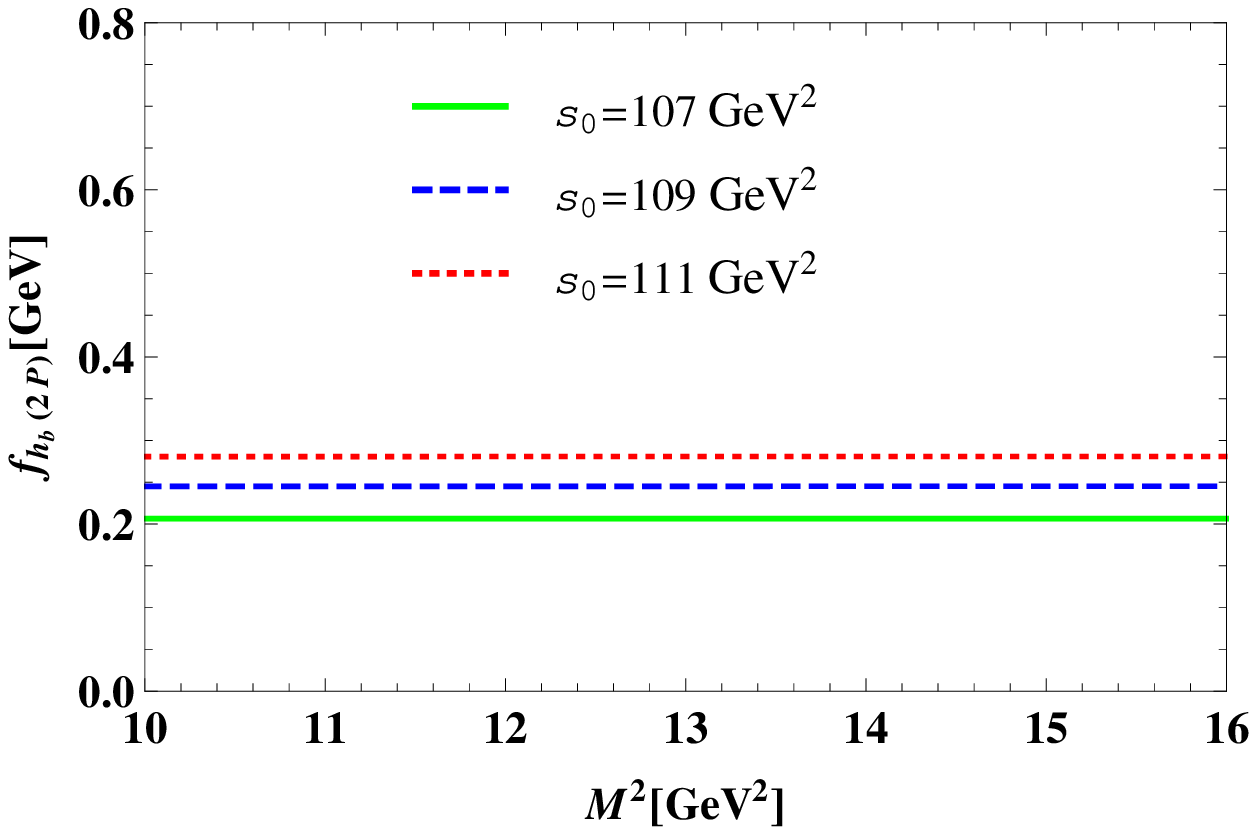}\,\,\,\,\,\,\,\,\,\,\,\,\,\,\,\,%
\includegraphics[totalheight=6cm,width=8cm]{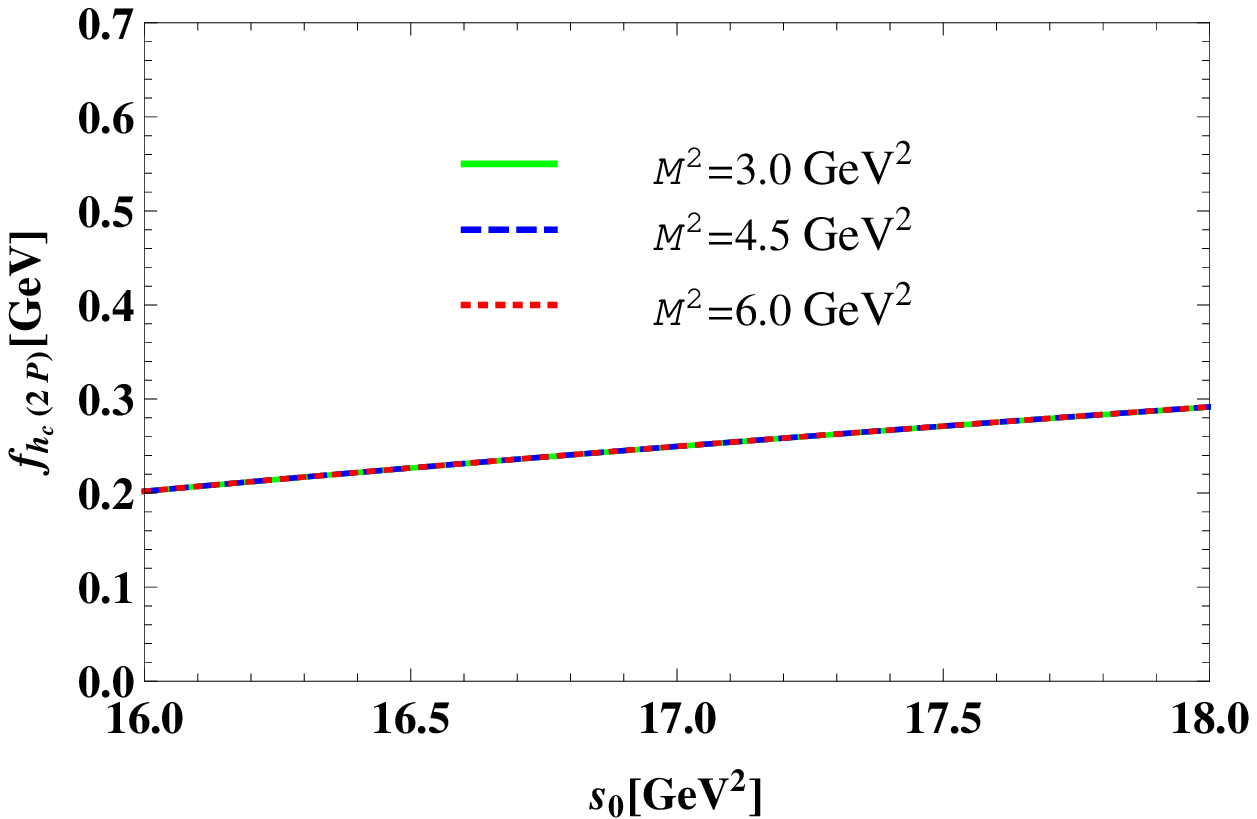}
\end{center}
\caption{ The same as figure 1 but for the decay constants $f_{h_c(1P,2P)}$.} \label{fig:DecCon1}
\end{figure}

\begin{figure}[h!]
\begin{center}
\includegraphics[totalheight=6cm,width=8cm]{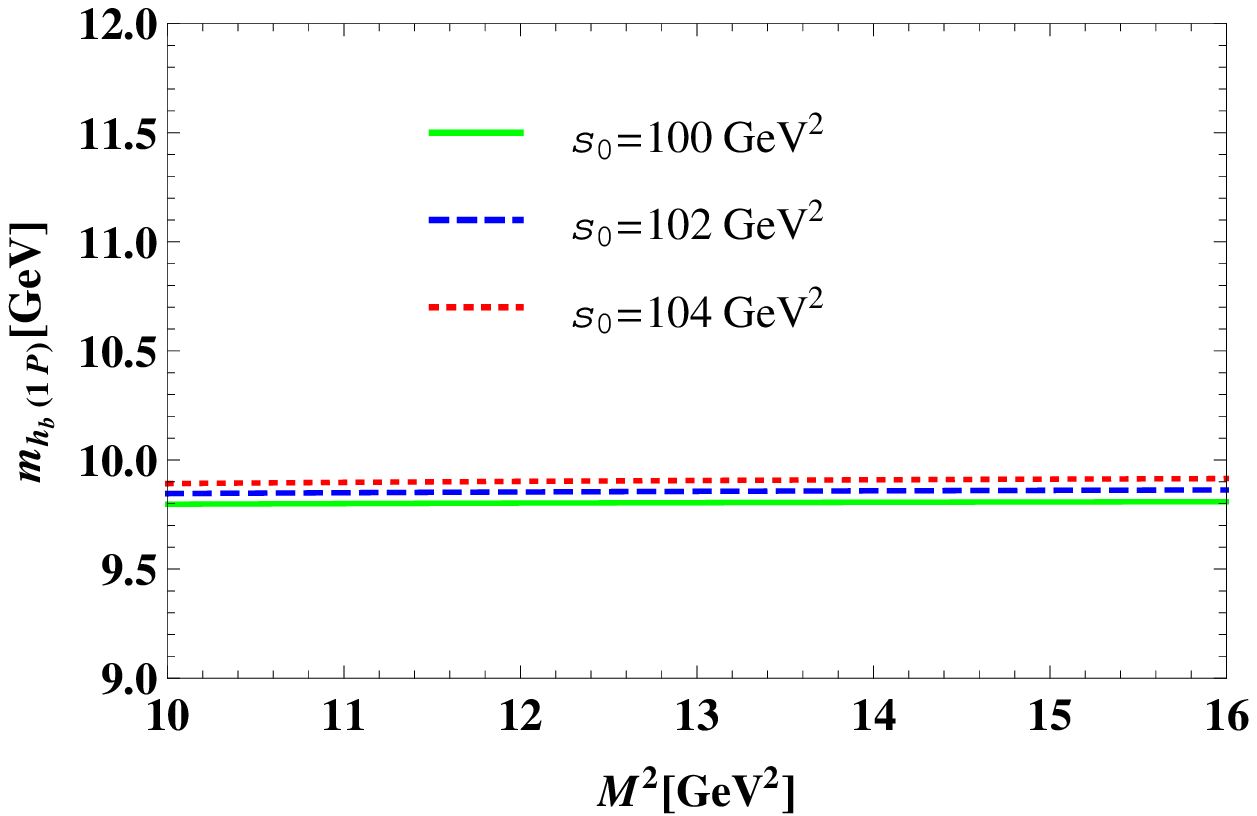}\,\,\,\,\,\,\,\,\,\,\,\,\,\,\,\,%
\includegraphics[totalheight=6cm,width=8cm]{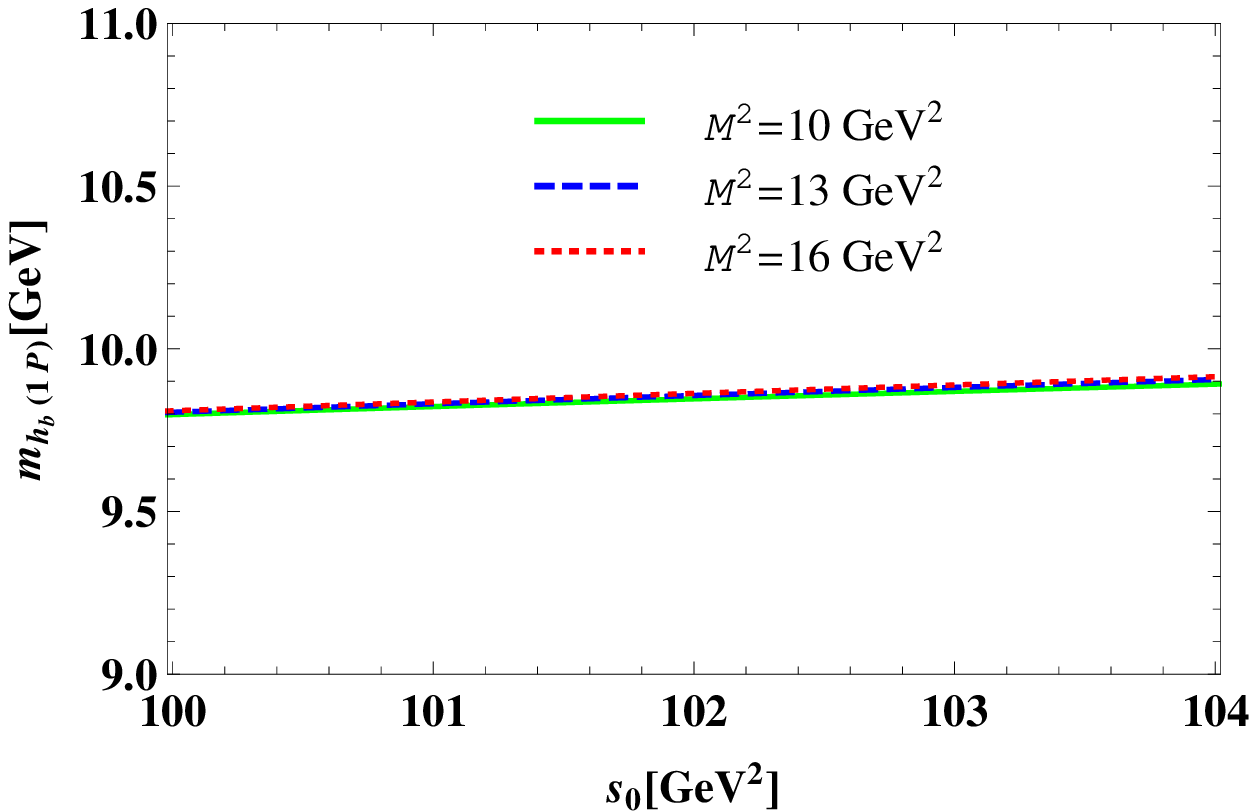}
\includegraphics[totalheight=6cm,width=8cm]{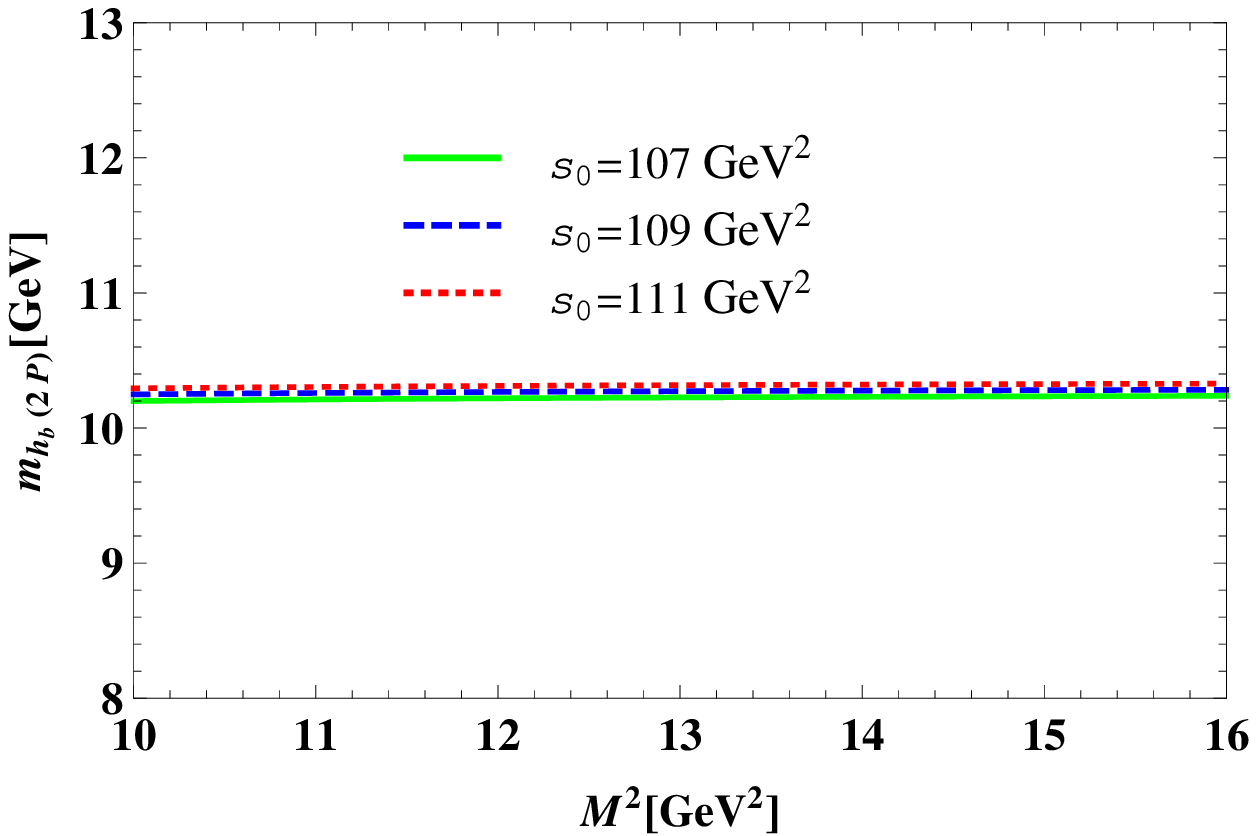}\,\,\,\,\,\,\,\,\,\,\,\,\,\,\,\,%
\includegraphics[totalheight=6cm,width=8cm]{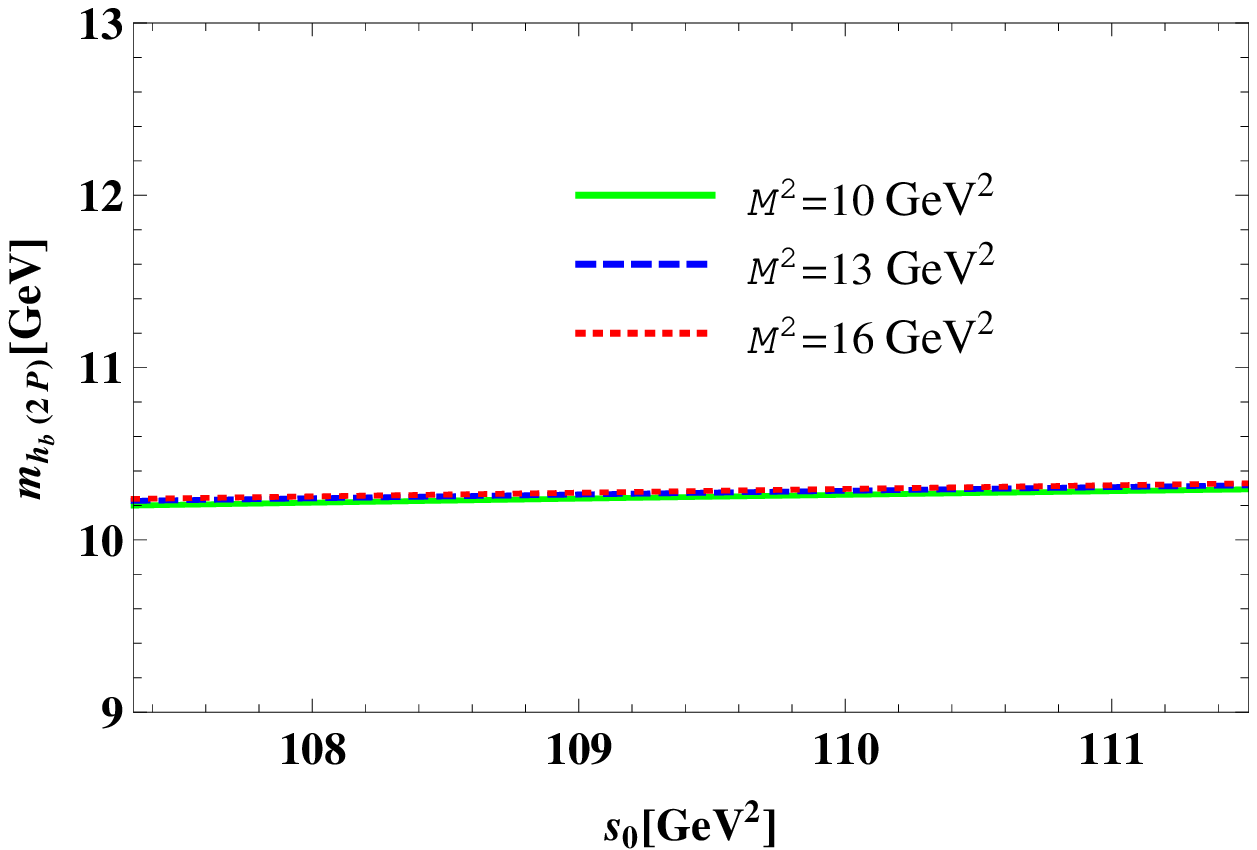}
\end{center}
\caption{ The same as figure 1 but for the masses of the mesons $h_{b}(1P,2P)$.} \label{fig:Masshb}
\end{figure}

\begin{figure}[h!]
\begin{center}
\includegraphics[totalheight=6cm,width=8cm]{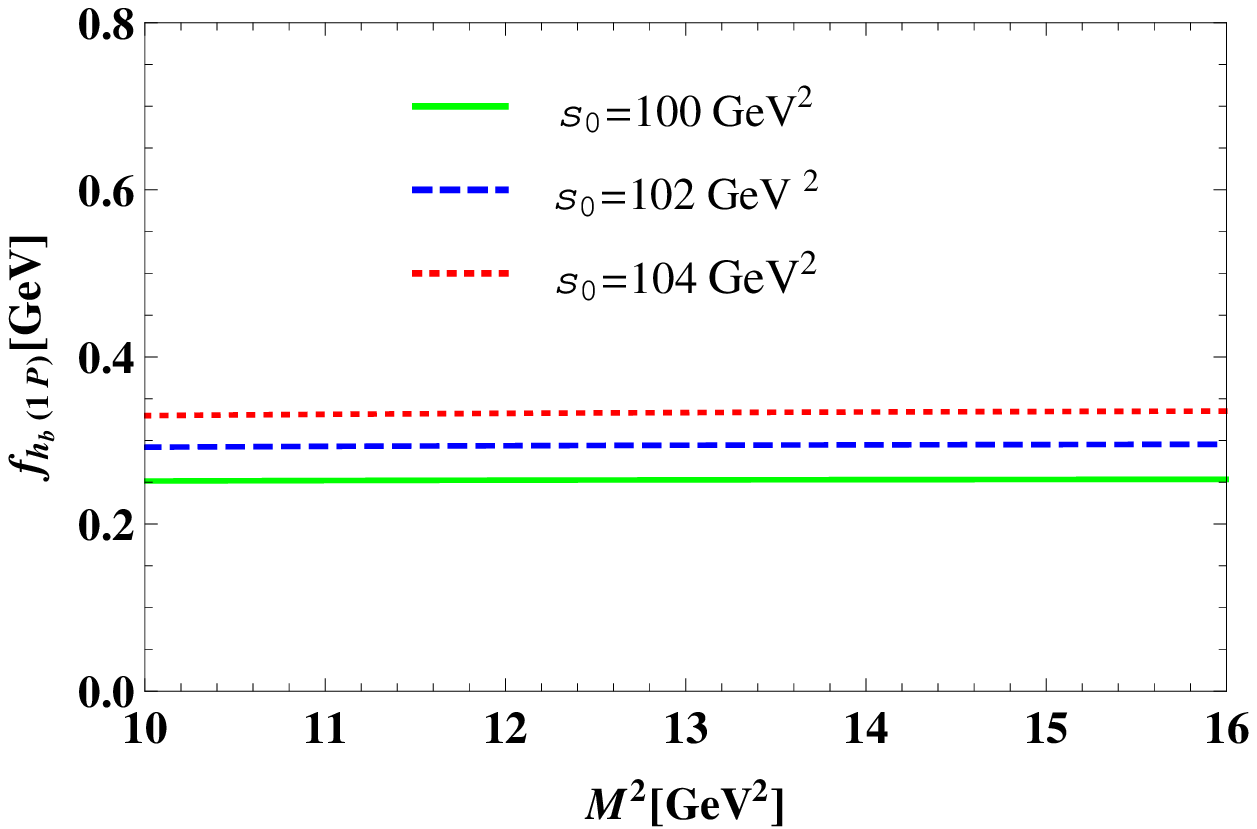}\,\,\,\,\,\,\,\,\,\,\,\,\,\,\,\,%
\includegraphics[totalheight=6cm,width=8cm]{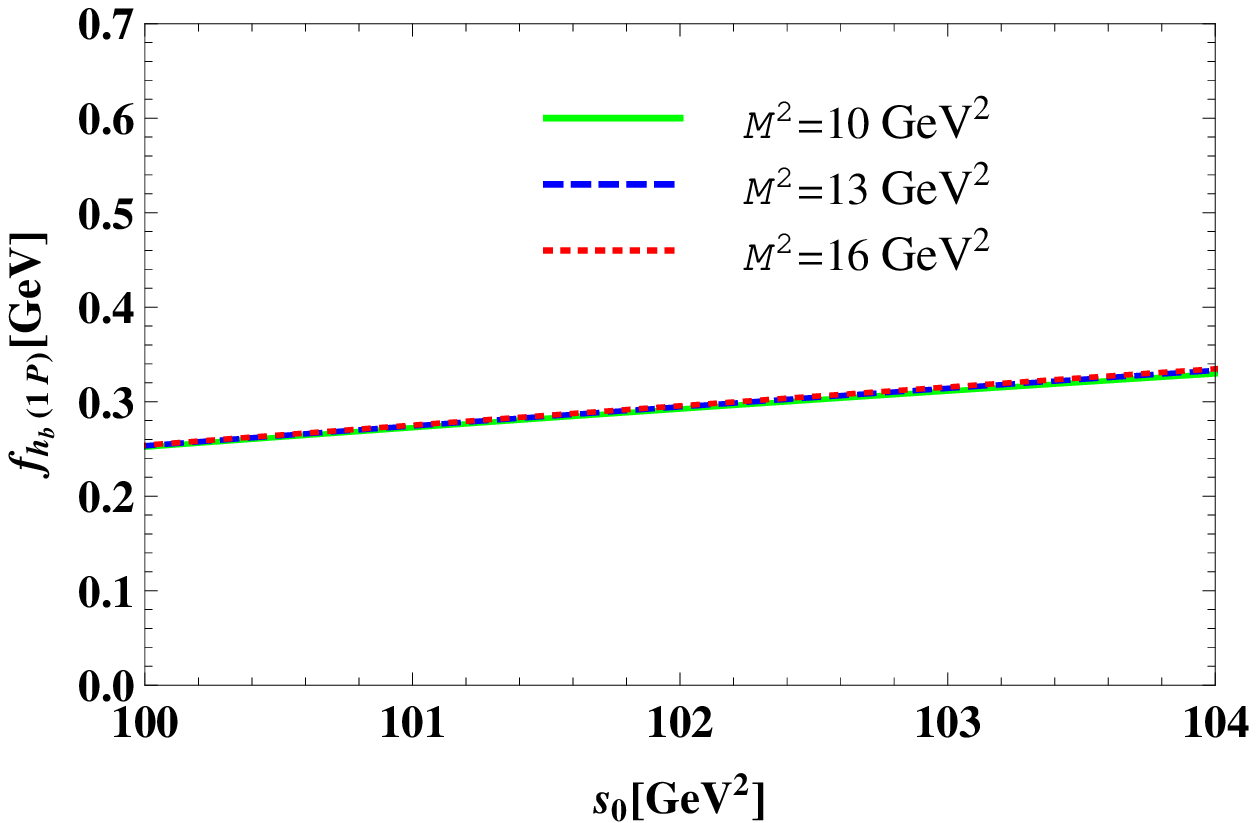}
\includegraphics[totalheight=6cm,width=8cm]{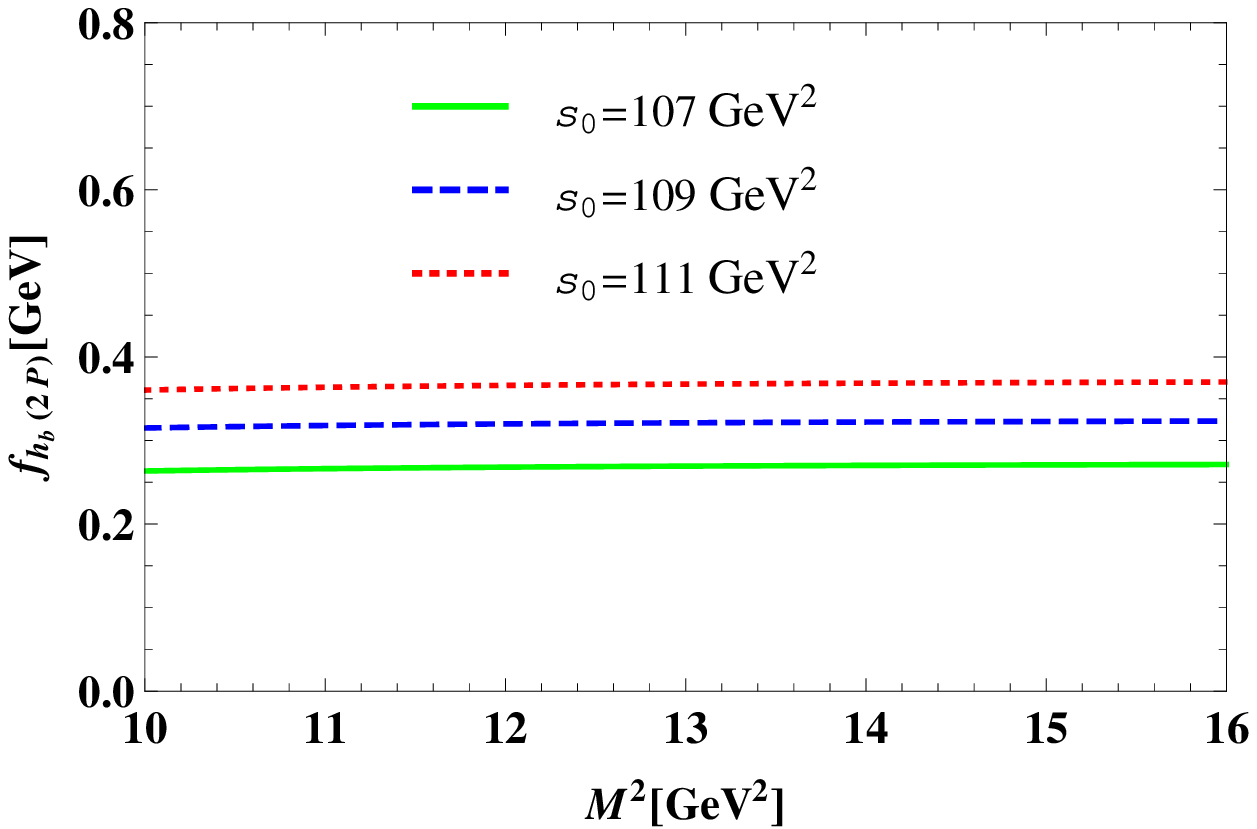}\,\,\,\,\,\,\,\,\,\,\,\,\,\,\,\,%
\includegraphics[totalheight=6cm,width=8cm]{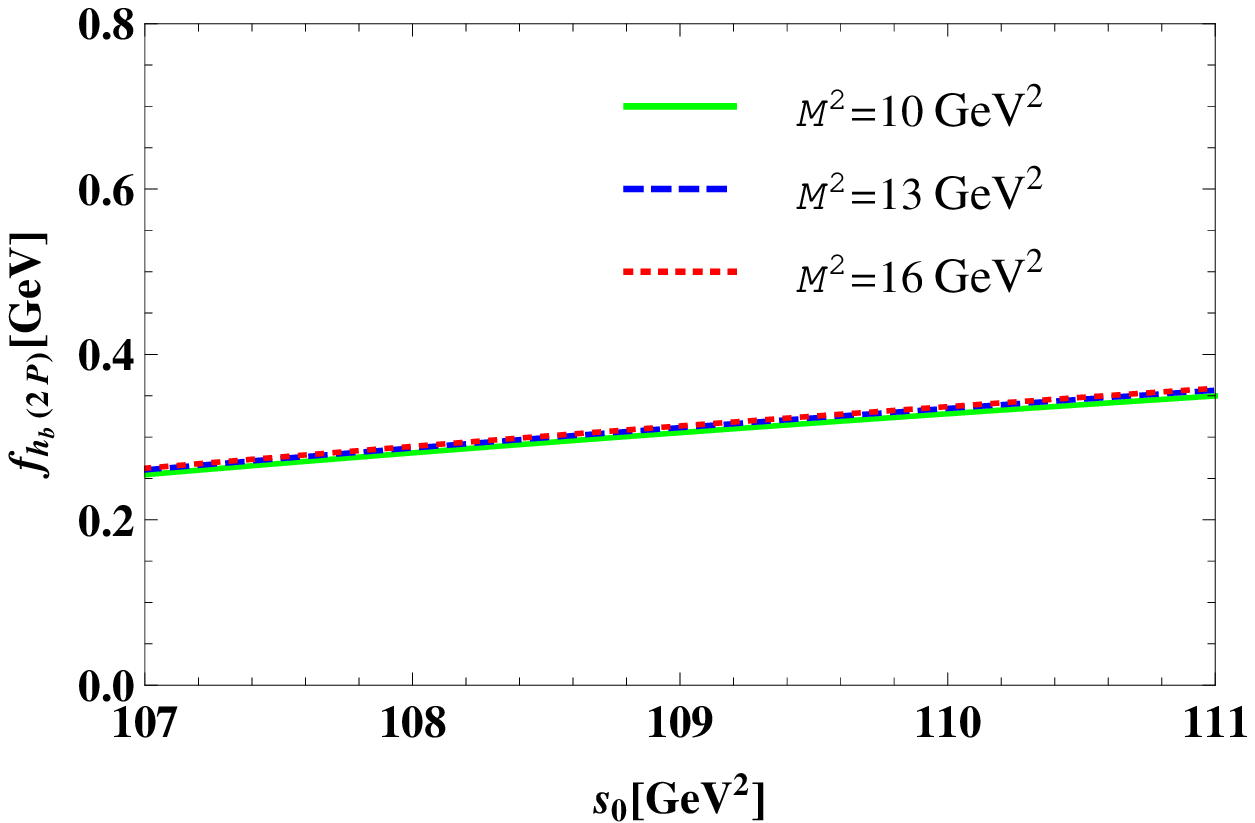}
\end{center}
\caption{ The same as figure 1 but for the decay constants $f_{h_{b}(1P,2P)}$.} \label{fig:DecConhb2P}
\end{figure}

\end{widetext}

\section{Concluding Remarks}

We have studied the
$h_{Q}(1P)$ and $h_{Q}(2P)$ systems
employing the QCD sum rule method, where in calculations terms up to
dimension eight have been used. We adopted an  interpolating current including a derivative and 
 with quantum numbers
$J^{PC} = 1^{+-}$ for the $h_{Q}(1P)$ and $h_{Q}(2P)$ states. The mass and decay constant of the ground-state
mesons $h_{Q}(1P)$  and their first radial excitations
$h_{Q}(2P)$ have been extracted from the corresponding sum rules
derived in the present study. Our results for the spectroscopic parameters of
these mesons, compared with the existing experimental data as well as other theoretical predictions, are presented in Tables\ \ref{tab:MassResults} and
\ref{tab:DecConResults}.

We should note that recently, in the framework of QCD sum rule
method, the mass and decay constant of the $h_b$ mesons were
calculated in  Ref.\ \cite{AgaevZb} using a tensor-type interpolating
current. For the mass and decay constant of the $h_b(1P)$ and
$h_b(2P)$ mesons in this work the following results were found:
for the ground-state $h_b(1P)$ meson
$m_{h_b}(1P)=9886^{+81}_{-78}\ \mathrm{MeV}$ and $f_{h_b} =
325^{+61}_{-57}\ \mathrm{MeV}$, for its first radial excited state
$m_{h_b}(2P)=10331^{+108}_{-117}\ \mathrm{MeV}$ and
$f_{h_b}(2P)=286^{+58}_{-53} \ \mathrm{MeV}$. Obtained in Ref.\
\cite{AgaevZb} by using the tensor-type current, the value of the decay constant for   $h_b(1P)$
is $9.85\%$ higher than,  and the value of the decay constant for the $h_b(2P)$ state is $10.06\%$
lower than the values extracted in the present work using the
axial-vector type current. Any future experimental data on the decay constants will help us to determine which interpolating current is favored for the states under consideration. As for the  masses,  however, the two studies
predict consistent values within the errors.  Our predictions on the masses
are also in  agreements with other theoretical predictions as
well as existing experimental data. There are considerable
differences among the theoretical predictions on the values of the
decay constants. The results of present work may shed light on
experimental searches for the $h_{c}(2P)$ state. Besides the predicted masses, the obtained decay constants can be used in theoretical determinations of the total widths of the considered states via analyses of their possible strong, electromagnetic and weak decays. Such theoretical predictions on the widths of these states may help experimental groups to measure the parameters of these states more precisely. 

The  determination of the basic properties of quarkonia is very important to explain  the experimental data exist on the spectrum of the hidden charmed and bottom sectors.  Such determinations will enable us to categorize the experimentally observed resonances and precisely determine which of these states belong to the quarkonia resonances and which ones to the class of the quarkonia-like XYZ exotic  states that are in agenda of the particle physicists nowadays. We
hope that the theoretical studies will improve our knowledge in
this regard, and shed light on the experiments in order to obtain more accurate data. 

\section*{ACKNOWLEDGEMENTS}

K.~A. thanks  Do\v{g}u\c{s} University for the partial financial
support through  the grant BAP 2015-16-D1-B04.  J. Y. S\"ung\"u appreciates the support of Kocaeli University through the grant BAP 2018/082.  The authors are
also grateful to S. S. Agaev and H. Sundu for useful discussions.

\appendix*

\section{The Non-perturbative Part of the Spectral Density}

\renewcommand{\theequation}{\Alph{section}.\arabic{equation}} \label{sec:App}

The non-perturbative part of the spectral density used in Eq.\
(\ref{eq:rho}) is found in terms of the
dimension four, six and eight gluon condensates as:
\begin{eqnarray}
\rho ^{\mathrm{Nonpert.}}(s)&=&\Big \langle\frac{\alpha
_{s}G^{2}}{\pi }\Big\rangle \int_{0}^{1}f_{1}(z,s)dz  \notag \\
&+&\Big \langle g_{s}^{3}G^{3}\Big \rangle\int_{0}^{1}f_{2}(z,s)dz  \notag \\
&+&\Big \langle\frac{\alpha _{s}G^{2}}{\pi }\Big \rangle^{2}%
\int_{0}^{1}f_{3}(z,s)dz.
\label{eq:NPert}
\end{eqnarray}
In Eq.\ (\ref{eq:NPert}) the functions $f_{1}(z,s),\ f_{2}(z,s)$
and $f_{3}(z,s)$ have the explicit forms:
\begin{eqnarray}
f_{1}(z,s)&=&\frac{1}{24r^2}\Big[-6r^2(s-\Phi)-3r[-2m_Q^2+sr)] \notag \\
&\times&\delta^{(1)}(s-\Phi)+ m_Q^2 s(1+2r)
\delta^{(2)}(s-\Phi)\Big], \notag \\
\end{eqnarray}
\begin{eqnarray}
f_{2}(z,s)&=&\frac{1}{15\cdot 2^{9}\pi^2 r^5}\Big\lbrace\delta^{(4)}(s-\Phi) \Big(2 m_Q^6[1+5 r(1 + r)] \notag \\
&-& m_Q^4r[7 + r (31 + 23r)]s+6 m_Q^2 s^2 r^3(1 + 2 r)\notag\\
&+& r^5 s^3\Big)+2\Big[12r^3 \delta^{(1)}(s-\Phi)[1+5 r(1 + r)] \notag \\
&+& 6sr^3 \delta^{(2)}[1+ r (7+11r)]+r\delta^{(3)}\Big(-3 m_Q^4 [1 \notag\\
&+& 5 r(1+ r)] + 9 m_Q^2 sr [1 + 2 r (2 + r)] + 2 s^2r^3 \notag\\
&\times&(2+ 7 r)\Big)\Big]\Big\rbrace,
\end{eqnarray}
and
\begin{eqnarray}
f_{3}(z,s)&=&\frac{1}{2^4\cdot3^3 r^{2}}
m_Q^2 \pi^2\Big[6r \delta^{(3)}(s-\Phi)+ \delta^{(5)}(s-\Phi) \notag \\
&\times& s (-m_Q^2 + r s)+2\delta^{(4)}(s-\Phi)[-m_Q^2 + s (1 + 3
r)]\Big], \notag \\
\end{eqnarray}
where we use the following notations
\begin{equation}
\ r=z(z-1),\ \ \Phi =\frac{m_{Q}^{2}}{z(1-z)}.
\end{equation}
In the above expressions
\begin{equation}
\delta^{(n)}(s-\Phi)=\frac{d^{n}}{ds^{n}}\delta (s-\Phi).
\end{equation}
%


\end{document}